\documentclass[amsmath,amssymb,amsfonts,aps,prb,twocolumn,superscriptaddress,nolongbibliography]{revtex4-2}

\usepackage{graphicx}
\usepackage{xcolor}
\usepackage{natbib,hyperref}

\newcommand{\bra}[1]{\left\langle#1\right|}
\newcommand{\ket}[1]{\left|#1\right\rangle}

\newcommand{\sbra}[1]{\langle#1|}
\newcommand{\sket}[1]{|#1\rangle}
\newcommand{\sbracket}[2]{\langle#1 | #2\rangle}

\newcommand{\p}{{\prime}}
\newcommand{\pp}{{\prime\prime}}

\newcommand{\up}{\uparrow}
\newcommand{\dn}{\downarrow}

\begin{document}


\title{Theory of intermolecular exchange in coupled spin-1/2 nanographenes}

\author{David Jacob}
\email{david.jacob@ehu.es}
\affiliation{Departamento de Pol\'{i}meros y Materiales Avanzados: F\'{i}sica, Qu\'{i}mica y Tecnolog\'{i}a, Universidad del Pa\'{i}s Vasco UPV/EHU,
  Av. Tolosa 72, E-20018 San Sebasti\'{a}n, Spain}
\affiliation{IKERBASQUE, Basque Foundation for Science, Plaza Euskadi 5, E-48009 Bilbao, Spain}

\author{Joaqu\'{i}n Fern\'{a}ndez-Rossier}
\affiliation{ International Iberian Nanotechnology Laboratory (INL), 4715-330 Braga, Portugal}
\altaffiliation[On leave from ]{Departamento de F\'{i}sica Aplicada, Universidad de Alicante, E-03690 San Vicente del Raspeig, Spain}
\date{\today}

\date{\today}

\begin{abstract}
Open-shell nanographenes can be covalently bonded and still preserve their local moments, forming interacting spin lattices. In the case of benzenoid nanographenes, the Ovchinnikov-Lieb rules anticipate the spin of the ground state of the superstructure and thereby the sign of the intermolecular exchange. Here we address the underlying microscopic mechanisms for intermolecular exchange in this type of system. We find that, in general, three different mechanisms contribute. First, Hund's ferromagnetic exchange that promotes ferromagnetic interactions of electrons in overlapping orbitals. Second, superexchange driven by intermolecular hybridization, identical to Anderson kinetic exchange, which is a decreasing function of the Hubbard-$U$ energy scale, and is always antiferromagnetic. Third, a Coulomb-driven superexchange, that increases as a function of $U$ and involves virtual excitation of excited molecular orbitals that are extended over the entire structure.  We find that Coulomb-driven superexchange can be either ferro- or antiferromagnetic, accounting for Ovchinnikov-Lieb rules. We compute these exchange energies for the case of coupled $S=1/2$ phenalenyl triangulenes, using multi configurational methods both with Hubbard and extended Hubbard models, thereby addressing the influence of long-range Coulomb interactions on the exchange interactions.
\end{abstract}

\maketitle


\section{Introduction}
\label{sec:intro}

The origin of intra-atomic ferromagnetic (FM) interactions and their connection to  exchange interactions was understood very early on,
together with the role of overlap of the participating quantum states~\cite{Hund}. The understanding of antiferromagnetic (AF) exchange
between distant atoms in magnetic insulators arrived much later~\cite{Anderson59} and highlighted the subtle interplay between
interatomic hybridizations and on-site Coulomb repulsion~\cite{auerbach,fazekas99}.

After many decades of theoretical work on open-shell magnetism in nanographenes (NGs)~\cite{longuet1950,borden77,ovchinnikov78,klein82,rajca94,fano98,JFR07,gucclu09,jung09,yazyev10,golor14,malrieu14,shi17,garciam17,ortiz18,ortiz19,jacob21}, relatively recent progress~\cite{mishra2019synthesis,li2020,zheng20,mishra2020,mishra2020c,ortiz20,mishra21a,mishra21b,hieulle2021,cheng22,de2022} in experimental techniques is now paving the way to its exploration in the laboratory. Unlike magnetism in conventional magnetic insulators, spin moments in NGs are hosted by non bonding molecular orbitals (MOs) formed by linear combinations of carbon $\pi$ orbitals. Thus  spin moments in $\pi$ magnetism are relatively extended, compared to the $d$ and $f$ electron counterparts~\cite{garciam17}. Importantly, experiments reporting magnetic excitations in {\em different} supramolecular structures formed by covalent bonding of {\em the same} molecular building blocks~\cite{mishra2020} calls for a rationalization of the collective spin properties in terms of the spin interactions between the parts.

Here we focus on the case of $S=1/2$ phenalenyl molecules as building blocks for larger structures. Phenalenyl is the smallest
member of the triangulene family~\cite{su2020}. Graphene triangulenes of various sizes are a prominent example of molecular
building blocks for $\pi$ magnetism in supramolecular structures~\cite{ortiz22}. For triangulenes with a lateral dimension of
$n$ benzenes ($[n]$triangulenes in the following), the spin $S$ of the ground state (GS) satisfies $2S=n-1$~\cite{JFR07}.
Thus phenalenyls are $[2]$triangulenes and their electronic ground state has $S=1/2$. On the other hand, $[3]$triangulene-based structures
have been reported~\cite{mishra2020,mishra21b,hieulle2021}, including dimers, trimers, tetramers, and rings, but also the triply
coordinated structures~\cite{cheng22} that may eventually lead to the formation of 2D crystals. There is strong  experimental
evidence that the open-shell nature of individual triangulenes is preserved in the triangulene arrays and electronic excitations
are compatible with strong intermolecular spin exchange~\cite{mishra2020,mishra21b}.

\begin{figure}
  \includegraphics[width=\linewidth]{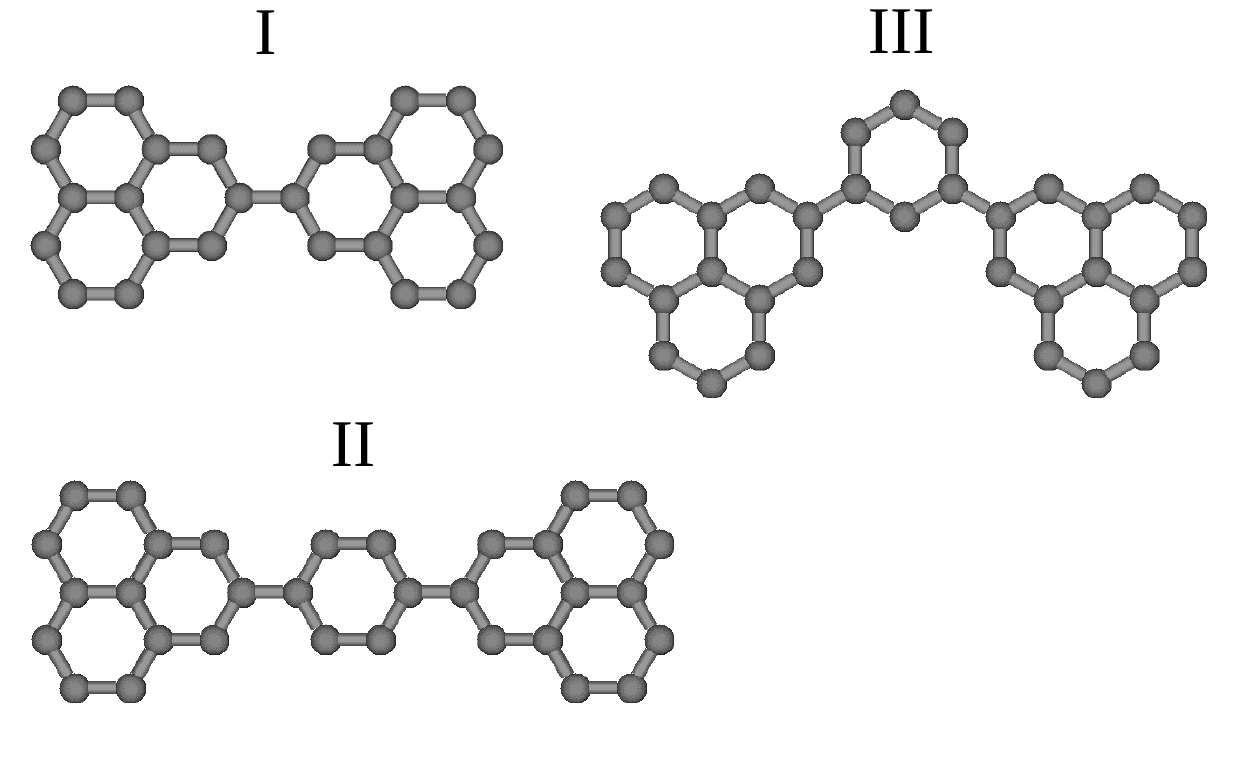}
  \caption{\label{fig:structures}
    Three different  phenalenyl dimer structures, I, II, III studied in this work:
    direct phenalenyl dimer (I); phenalenyl dimer with a $1,4$-phenylene spacer;
    phenalenyl dimer with a $1,3$-phenylene spacer (III). Structures I and II are
    sublattice balanced, whereas III has $N_A=N_B+2$.
  }
\end{figure}

The spin of the GS of a given benzenoid molecule can be anticipated from the Ovchinnikov rule~\cite{ovchinnikov78} that can be
upgraded to Lieb's theorem~\cite{Lieb89}, if we model the molecules with a Hubbard Hamiltonian~\cite{JFR07}: we can breakdown
the carbon sites in two interpenetrating triangular lattices, $A$ and $B$, count the number of carbon atoms $N_A$ and $N_B$ in each
class, and predict the spin to be $2S=|N_A-N_B|$. This rule naturally accounts for the spin of triangulenes, as magnetic building
blocks, and also molecular structures formed with them. Thus the phenalenyl dimers I and  II  (shown  in Fig.~\ref{fig:structures})
are expected to have $S=0$ GSs whereas dimer III, in the same figure, is expected to have $S=1$. Since the molecule remains open shell
after covalent bonding, the spin of their GS automatically gives us the sign of the intermolecular exchange. The goal of this paper is
to provide a general theory for understanding the mechanisms of intermolecular spin couplings between phenalenyls, although our theory
can be easily extended to other radical NGs, and the main principles apply to larger triangulenes.

We find three types of exchange interactions. First, Hund's FM coupling for electrons in MOs that have some overlap in real
space. This mechanism is already present at the intramolecular level~\cite{ortiz19} and accounts for the linear scaling of the spin of
the ground state and the size of triangulenes~\cite{JFR07,gucclu09}  or extended triangulenes~\cite{li2020,jacob21}, but it may also
happen for coupled triangulenes when the coupling occurs indirectly via a third molecule (as e.g. a single benzene)~\cite{cheng22}.
Hund's FM coupling arises from the direct exchange matrix elements of the Coulomb interaction, which ultimately are also
responsible for Hund's rule magnetism in open-shell atoms~\cite{Hund}. 

Second, we have two types of superexchange interactions. Very much like superexchange in atomic insulators, intermolecular exchange (IE)
depends on the interplay between two competing interactions: on the one hand, intermolecular hybridizations of the MOs hosting the spins
promotes electron delocalization and double occupancy. On the other hand, Coulomb repulsion opposes those processes. We refer to this as
{\em kinetic IE}. It is always AF. We note that, in some cases, such as triangulenes and the Clar's goblet,
  a nanographene diradical with bowtie shape~\cite{clar72},
intermolecular first-neighbor covalent bonding does not lead to intermolecular hybridization of the non bonding MOs that host the spins. In these
cases, kinetic IE is governed by third neighbor hopping, $t_3$.
Delocalization can also be promoted if the electrons occupy otherwise empty molecular orbitals that are extended across different molecules.
This virtual occupation is driven by quasi particle scattering generated by  Coulomb interactions. We refer to this as {\em Coulomb-driven IE}.
This type of IE can be both AF or FM. In general, all mechanisms can coexist.

The rest of this work is organized as follows. In Sec.~\ref{sec:model} we introduce the model
Hamiltonian used for the description of NGs. In Sec.~\ref{sec:sp_states}
we discuss the single-particle spectra of these model Hamiltonians for the three different
structures shown in Fig.~\ref{fig:structures}. In Sec.~\ref{sec:cas} we describe the complete
active space method used for the solution of the many-body problem.
In Sec.~\ref{sec:ie_mechanisms} we discuss and analyze in detail the different exchange mechanisms active
in open-shell NGs. A special focus is the Coulomb-driven superexchange mechanism for which we
derive an analytic expression in second-order perturbation theory (Sec.~\ref{sub:cdse}).
In Sec.~\ref{sec:numeric} we discuss and analyze numerical results for the AF dimers I and II
(Sec.~\ref{sub:af_dimers}) as well as the FM case of dimer III (Sec.~\ref{sub:fm_dimer}).
In Sec.~\ref{sec:full_pt} we generalize and apply the perturbation theory for Coulomb-driven superexchange
to the full single-particle spectrum. 
In Sec.~\ref{sec:lr_coulomb} we study numerically and analytically the effect of long-range Coulomb interactions
on the different exchange mechanisms.
Finally, in Sec.~\ref{sec:conclusions} we wrap-up and present our main conclusions.

\section{Model}
\label{sec:model}

We consider an extended Hubbard model as an effective description for the $\pi$ orbitals
of the carbon atoms in a NG, taking into account hopping between first
and third neighbors, as well as non local parts of the Coulomb interaction in addition
to the local Hubbard-type interaction:
\begin{equation}
  \label{eq:H}
  \mathcal{H} = \mathcal{H}_{\rm sp} + \mathcal{H}_U + \mathcal{H}_V + \mathcal{H}_K
\end{equation}
The first term in (\ref{eq:H}) is the single-particle part describing hopping of electrons between different carbon lattice sites: 
\begin{equation}
  \label{eq:Ht}
 \mathcal{H}_{\rm sp} =  \mathcal{H}_t + \mathcal{H}_{t_3} 
  = -t \sum_{\langle i,j \rangle, \sigma} c_{i\sigma}^\dagger c_{j\sigma}
  -t_3 \sum_{\langle\langle\langle i,j \rangle\rangle\rangle,\sigma} c_{i\sigma}^\dagger c_{j\sigma}
\end{equation}
where $c_{i\sigma}^\dagger$ ($c_{i\sigma}$) creates (destroys) an electron of spin
$\sigma\in\{\up,\dn\}$ at carbon site $i$, and $t$ and $t_3$ are the amplitudes for first and third nearest neighbor
hopping between lattice sites, respectively. $\langle i,j \rangle$ and $\langle\langle\langle i,j \rangle\rangle\rangle$ indicate
restriction of sites $i$ and $j$ to first and third nearest neighbors, respectively. 
The second neighbor hopping is usually neglected in this approach because it
breaks electron-hole symmetry, leading to an inhomogeneous distribution of the charge, that is counterbalanced by
long-range Coulomb interactions. The second term in (\ref{eq:H}) is the Hubbard interaction describing repulsion $\sim{U}$
between electrons on the same site:
\begin{equation}
  \label{eq:Hubbard}
  \mathcal{H}_U = U \sum_i \hat{n}_{i\up} \hat{n}_{i\dn}  
\end{equation}
$\hat{n}_{i\sigma}=c_{i\sigma}^\dagger c_{i\sigma}$ is the occupation number operator for carbon site $i$ and spin $\sigma$.
We note that the Hubbard model described by the first two terms already yields an excellent description for the
low-energy physics (e.g. Kondo physics and spin excitations) of open-shell NGs for $t=2.7$~eV and $t_3=0.1t$ and
for $U$ in the range $|t|\le U \le 3|t|$~\cite{ortiz19,jacob21}.

In order to investigate the effect of the long-range (LR) part of the Coulomb interaction,
we will occasionally also include the third and fourth term in (\ref{eq:H}), describing 
(i) Coulomb repulsion between electrons at different carbon sites,
\begin{equation}
  \label{eq:Coul}
  \mathcal{H}_V =  \sum_{i<j} V_{i,j} \, \hat{n}_i \hat{n}_j
\end{equation}
and (ii) direct Coulomb exchange,
\begin{equation}
  \label{eq:Exch}
  \mathcal{H}_K = \sum_{i<j,\sigma,\sigma^\prime} K_{i,j} \, c_{i\sigma}^\dagger c_{j\sigma^\prime}^\dagger c_{i\sigma^\prime} c_{j\sigma}
\end{equation}
Note that the direct exchange term gives rise to Hund-type coupling $\sim -K_{i,j} \, \vec{S}_i \cdot \vec{S}_j$  between
electron spins $\vec{S}_i=\sum_{\sigma\sigma^\prime} c_{i\sigma}^\dagger \, \vec\tau_{\sigma\sigma^\prime} \, c_{i\sigma^\prime}$ and
$\vec{S}_j=\sum_{\sigma\sigma^\prime} c_{j\sigma}^\dagger \, \vec\tau_{\sigma\sigma^\prime} \, c_{j\sigma^\prime}$ on different carbon sites,
which is FM in nature ($K_{i,j}>0$).

In order to estimate the LR Coulomb interaction parameters $V_{i,j}$ and $K_{i,j}$ we have computed the
bare Coulomb matrix for the $\pi$ orbitals of the carbon atoms in the idealized geometries shown in Fig.~\ref{fig:structures}
using the quantum chemistry code Gaussian09~\cite{G09}, as described in more detail in Sec.~\ref{sec:lr_coulomb} and the Appendix.

\section{Single-particle states}
\label{sec:sp_states}

We briefly discuss the properties of the single-particle states that are most relevant for understanding intermolecular exchange.
The single-particle states are the molecular orbitals (MOs), obtained by diagonalizing the one-body (hopping) part of (\ref{eq:H}),
i.e., $\mathcal{H}_t \sket{\psi_k} = \epsilon_k \, \sket{\psi_k}$ and 
$\sket{\psi_k} = \sum_{i\in{\rm sites}} \psi_k(i) \sket{i}$ where the expansion coefficient
$\psi_k(i)\equiv\sbracket{\psi_k}{i}$ is the ``wave function'' of the MO in the site basis.

Because of the bipartite nature of the single-particle model~\cite{sutherland86,JFR07,pereira08,ortiz19} that endows the Hamiltonian
with chiral symmetry, the single-particle spectra have three types of states, $\epsilon_k>0$, $\epsilon_k<0$ and $\epsilon_k=0$.
Finite-energy states come in electron-hole symmetric pairs, whose wave functions are related due to the chirality of the single-particle
Hamiltonian~\cite{pereira08,ortiz19}. On the other hand, the zero-energy modes or simply zero modes (ZMs) with $\epsilon_k=0$
are located inside a gap of order $t$. The ZMs are usually localized at the borders of the NG (edge states) and at charge neutrality are
singly occupied, thus giving rise to an open shell from which the magnetism of NGs is derived.

\begin{figure}
  \includegraphics[width=0.99\linewidth]{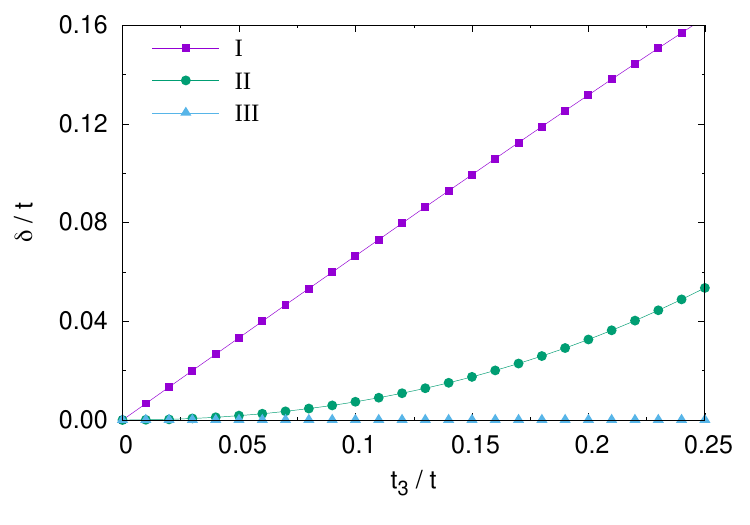}
  \caption{\label{fig:deltasp}
    Single-particle splitting $\delta$ of the ZMs of structures I, II and III as a function of the third-neighbor hopping $t_3$.
 }
\end{figure}

Even for $t_3=0$ the wave functions of finite energy MOs have weight on both triangulenes and both sublatttices.
In contrast, for $t_3=0$ dimers I and II have two ZMs that, on account of the lack of global sublattice polarization,
can be chosen to be sublattice polarized~\cite{ortiz19} (note that here and below ``polarization'' and
  ``polarized'' do not denote spin polarization but an imbalance in the number of sublattice sites, i.e., $N_A\neq N_B$).
With this choice, the two  ZMs are localized in just one triangulene and one sublattice.  
For dimer III, the two ZMs are polarized in the same sublattice and, because of the  the  mirror symmetry of the molecule,
we can find a representation where each ZM is localized in a different phenalenyl.
As we switch on $t_3$, the ZMs of structures
I and II  split and the  resulting molecular orbitals become linear combinations of the two ZMs~\cite{ortiz22}.
As we show in Fig.~\ref{fig:deltasp} the splitting of the ZMs, $\delta$, scales linearly (quadratically)
with $t_3$ for structure I (II). Specifically, the single-particle splittings of structures I and II are given by
\begin{equation}
  \label{eq:deltasp}
  \delta_{\rm I} = 2\cdot\frac{t_3}{3} \hspace{1ex} \mbox{ and } \hspace{1ex} \delta_{\rm II} = 0.72\cdot\frac{t_3^2}{t},
\end{equation}
respectively. The  equation for $\delta_{\rm I}$ can be derived analytically~\cite{ortiz22}, whereas the result for $\delta_{\rm II}$
has been obtained numerically. For dimer I,  the two ZMs have weight in sites that are third neighbors, so that intermolecular
hybridization of ZMs is enabled directly by $t_3$.
In contrast, for dimer II, the ZMs are farther away, so that intermolecular hybridization requires the participation of non-zero modes via
second order coupling.

In contrast,  for strucure III, the two ZMs keep $\epsilon_k=0$, ensured by the global sublattice imbalance
of this structure~\cite{JFR07,ortiz19}. They are both located in the majority sublattice
(note that``majority'' does not refer to the majority spin but to the sublattice type with the larger number of sites).
The wave functions of ZMs and the MOs closest in energy are shown in the bottom panels of Fig.~2. 

For later convenience, we introduce the following notation for the localized ZMs:
\begin{equation}
  \ket{z_\zeta} = \sum_i z_\zeta(i) \ket{i} 
\end{equation}
where $\zeta=1,2$ denotes the ZM localized in one of the triangulenes.
Furthermore, positive and negative energy states $\ket{\psi_k}$ will be labeled by positive and negative integers, respectively,
i.e., $k\in\{\pm1,\pm2,\ldots\}$.

\begin{figure*}
  \includegraphics[width=0.9\linewidth]{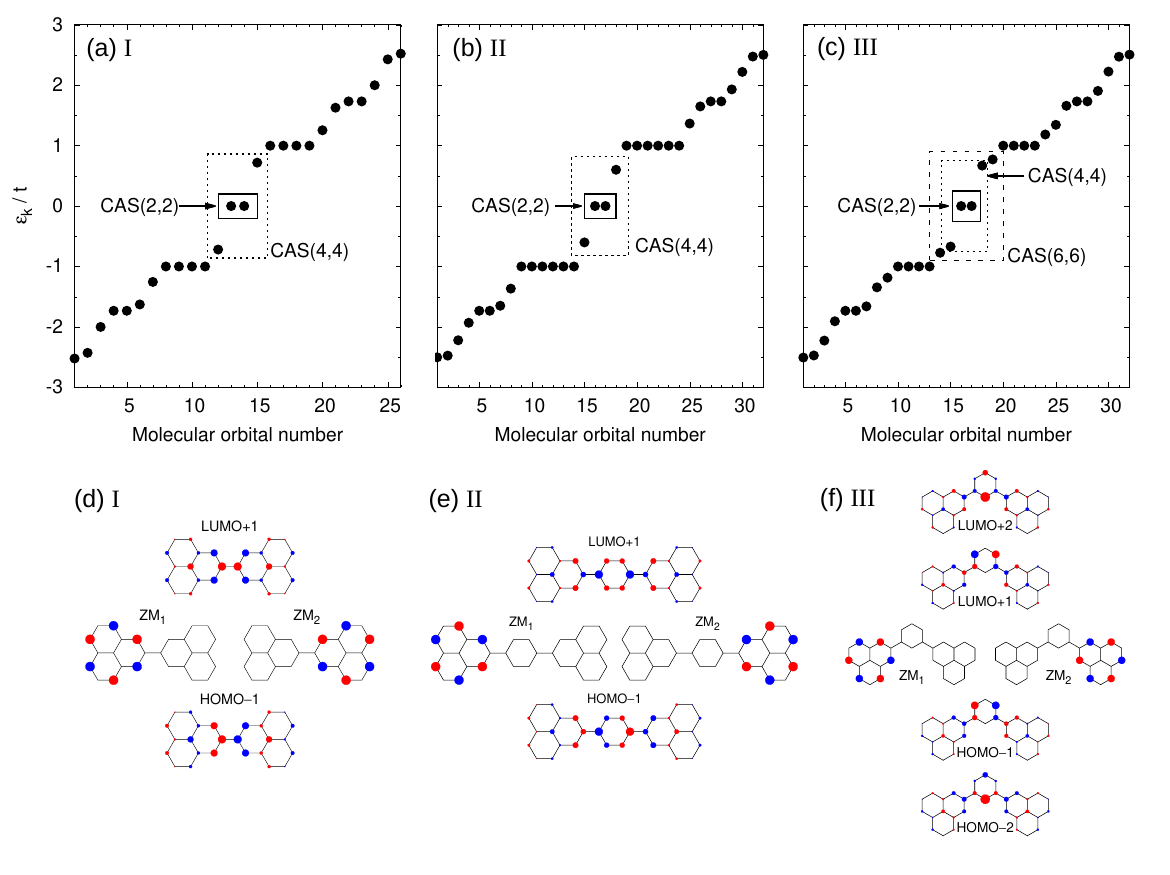}
  \caption{\label{fig:singleparticle}
    Single-particle spectra for structures I, II, and III [panels (a), (b) and (c), respectively] calculated for $t_3=0$
    so that all three spectra have two zero-energy states. The corresponding MOs included in the CAS($N$,$N$)
    calculations are shown in panels (d), (e) and (f). The degenerate ZMs are taken so that they are fully localized in a single phenalenyl.
 }
\end{figure*}

\section{Complete active space}
\label{sec:cas}

Due to the exponential growth of the many-body Hilbert space with the number of carbon sites, the Hamiltonian
(\ref{eq:H}) cannot be diagonalized numerically for more than a few sites ($<12$). For the numerical calculations
we therefore restrict the underlying one-body Hilbert space to a small subset C of MOs around the ZMs,
as shown in Fig.~\ref{fig:singleparticle} for the three systems considered here.

According to  Lieb's theorem the number of ZMs is precisely given by the sublattice imbalance $|N_A-N_B|$
of the NG. Additionally, we may also include the MOs closest in energy to the ZMs, i.e., the level(s) below
the highest occupied orbital(s), HOMO-1 and HOMO-2, and the level(s) above the lowest unoccupied orbital(s),
LUMO+1 and LUMO+2, as shown in Fig.~2.

Projecting the many-body Hamiltonian onto the restricted subspace then yields:
\begin{equation}
  \label{eq:HC}
  \mathcal{H}_{\rm C} = \sum_{k\in{\rm C}} \epsilon_k \, \hat{N}_k +
  \frac{1}{2}\sum_{ {{k,k^\p} \atop {q,q^\p}} \atop {\sigma,\sigma^\p}} \mathcal{W}_{k k^\p q q^\p} \, C_{k\sigma}^\dagger C_{k^\p\sigma^\p}^\dagger  C_{q^\p\sigma^\p} C_{q\sigma}
\end{equation}
where $C_{k\sigma}^\dagger$ ($C_{k\sigma}$) creates (destroys) one electron of spin $\sigma$ in MO $\psi_k$, $N_k=\sum_\sigma C_{k\sigma}^\dagger C_{k\sigma}$
measures the total occupation of MO $\psi_k$, and $\mathcal{W}_{k k^\p q q^\p}=\bra{\psi_k,\psi_{k^\p}}\hat{W}\ket{\psi_q,\psi_{q^\p}}$ are the
matrix elements of the total Coulomb interaction $\hat{W}\equiv\mathcal{H}_U+\mathcal{H}_V+\mathcal{H}_K$ in the MO basis,
$\mathcal{W}_{k k^\p q q^\p}=\mathcal{U}_{k k^\p q q^\p}+\mathcal{V}_{k k^\p q q^\p}+\mathcal{K}_{k k^\p q q^\p}$.
The Hubbard interaction part of the matrix is given by  
\begin{eqnarray}
  \label{eq:Umat}
  \mathcal{U}_{k k^\p q q^\p} &=& \bra{\psi_k,\psi_{k^\p}}\mathcal{H}_U\ket{\psi_q,\psi_{q^\p}} \nonumber\\
  &=& U \sum_{i\in{\rm sites}} \psi_k^\ast(i) \psi_{k^\p}^\ast(i) \psi_q(i) \psi_{q^\p}(i)
\end{eqnarray}
Note that in the MO basis the Hubbard interaction becomes generally ``non-local'' (i.e. interaction between different MOs),
even though it is local in the site basis.
Furthermore, for the matrix elements of the LR part of the Coulomb repulsion in the MO basis we obtain 
\begin{eqnarray}
  \mathcal{V}_{k k^\p q q^\p} &=& \bra{\psi_k,\psi_{k^\p}}\mathcal{H}_V\ket{\psi_q,\psi_{q^\p}} \nonumber\\
  &=& \sum_{{i,j\in{\rm sites}}\atop{i<j}}  V_{i,j} \, \psi_k^\ast(i) \psi_{k^\p}^\ast(j) \psi_q(i) \psi_{q^\p}(j)
\end{eqnarray}
while the direct Coulomb exchange matrix elements in the MO basis are given by:
\begin{eqnarray}
  \mathcal{K}_{k k^\p q q^\p} &=& \bra{\psi_k,\psi_{k^\p}}\mathcal{H}_J\ket{\psi_q,\psi_{q^\p}} \nonumber\\
  &=& \sum_{{i,j\in{\rm sites}}\atop{i<j}}  K_{i,j} \, \psi_k^\ast(i) \psi_{k^\p}^\ast(j) \psi_q(j) \psi_{q^\p}(i)
\end{eqnarray}

The projected Hamiltonian (\ref{eq:HC}) can be diagonalized for a fixed number of electrons $N_e$.
The dimension of the resulting restricted many-body Hilbert space, also called complete active space (CAS),
is completely determined by the number $N_{\rm C}$ of MOs making up the restricted one-body subspace C and
the number of electrons $N_e$, denoted as CAS($N_{\rm C}$,$N_e$). Here we consider always
charge-neutral species, described by a half-filled Hubbard model. Choosing the subspace C to be symmetric
around the ZMs requires $N\equiv{N_e=N_{\rm C}}$ at half-filling.

\section{Intermolecular exchange  mechanisms}
\label{sec:ie_mechanisms}

Structures I, II and III are all formed by
two identical radical NG molecules.  Depending on the bonding geometry,  they can be either sublattice balanced  (I,II)
or imbalanced (III). According to the Ovchinikov-Lieb (OL) rule~\cite{ovchinnikov78,Lieb89}, this will determine the spin of the ground state of the dimer.
Expectedly, our calculations with  the Hubbard model are in agreement with the OL rule and, for the systems considered here, predict
an open-shell structure for the dimer.  Consequently, the resulting low energy many-body states can be rationalized in terms
of an effective spin model:
\begin{equation}
  {\cal H}_{\rm eff}=  J\,\vec{S}_1\cdot\vec{S}_2
  \label{eq:Heis}
\end{equation} 
where the  intermolecular exchange  $J$ can be either FM or AF.  As we show now, there are three types of
intermolecular exchange:

\begin{enumerate}
\item Hund exchange, that is always FM and scales linearly with the strength of the Coulomb interaction. 
\item Kinetic superexchange, that scales quadratically  with the intermolecular hybridization and is inversely proportional to the
  strength of the Coulomb interaction.  It is always AF. 
\item Coulomb-driven superexchange, that can take both signs, and scales quadratically with the strength of the Coulomb interaction
\end{enumerate}

To a very good approximation, these three contributions are additive  in the relevant range of values for hopping and Coulomb
interactions, so that the exchange $J$ in Eq.~(\ref{eq:Heis}) can be written as the sum of three terms:
\begin{equation}
  \label{eq:Jeff}
  J\simeq J_{\rm Hund} + J_{\rm kin} + J_{\rm cd}
\end{equation}
where the three terms on the right hand side stand for Hund exchange, kinetic superexchange, and Coulomb-driven superexchange, respectively.

\subsection{Hund's exchange}
\label{sub:hund}

This interaction arises from the reduction of the Coulomb repulsion of electrons that overlap in space and have symmetric spin wave functions.
It is the same mechanism responsible of Hund rule's in atoms, that favor large spin for open shell configurations. For the three cases considered
here, it is only relevant for system III when $t_3>0$, as that is the only case where wave functions of the two ZMs that host the unpaired electrons
overlap. The theory of Hund's exchange for NG diradicals, projecting a Hubbard model into the basis of ZMs, was worked
out by one of us elsewhere~\cite{ortiz19}. In terms of the Hubbard matrix elements of Eq.~(\ref{eq:Umat}), we can write:
\begin{equation}
  J_{\rm Hund}=  -\mathcal{U}_{1221}
\end{equation}
where $1$ and $2$ label the ZMs localized on the phenalenyls $1$ and $2$. On the other hand, for the systems I and II,
$J_{\rm Hund}=0$ in the Hubbard model. 

\subsection{Kinetic superexchange}
\label{sub:kin_exch}

The first superexchange mechanism is the well known Anderson exchange mechanism~\cite{Anderson59}. It only involves the
ZM orbitals that host the unpaired electrons. Thus, for radical dimers there are two ZMs. The representation of the
Hubbard Hamiltonian in the basis of these two ZMs maps onto an effective Hubbard model for a dimer~\cite{ortiz18}.

If we take the third-neighbor hopping $\mathcal{H}_{t_3}$ as a perturbation to the tight-binding Hamiltonian $\mathcal{H}_{t}$,
the effective hopping $\tau$ between the two ZMs is given up to second order by
\begin{equation}
  \label{eq:tau}
  \tau = \bra{z_1} \mathcal{H}_{t_3} \ket{z_2}
  - \sum_{k_\pm} \frac{\bra{z_1} \mathcal{H}_{t_3} \ket{k_\pm}\bra{k_\pm} \mathcal{H}_{t_3} \ket{z_2}}{\epsilon_{k_\pm}}
\end{equation}
For dimer I the first order term linear in $t_3$ dominates. For dimer II, on the other hand, the first term is zero, since the ZMs are
too far away from each other to directly connect via third-neighbor hopping. Thus, only the second-order term quadratic in $t_3$ contributes
in this case.

The intermolecular hybridization $\tau$ lifts the degeneracy of the ZMs, resulting in two molecular orbitals that are
linear combinations of the ZMs of the building blocks with energies $\pm\tau$, and thus an energy splitting of $\delta=2\tau$. 
Explicit expressions for $\delta$ are given by Eq.~(\ref{eq:deltasp}) for dimers I and II.

The effective  interaction of the Hubbard dimer, $\cal U$, is given by
\begin{equation}
  \mathcal{U} \equiv \bra{z_\zeta,z_\zeta} \mathcal{H}_U \ket{z_\zeta,z_\zeta} = U \sum_i |z_\zeta(i)|^4 
  \label{eq:Ueff}
\end{equation}
where $\zeta=1,2$ labels the ZMs of the building blocks.

In the cases considered here we find ${\cal U}\gg\tau$ so that we can use the well established mapping
of the Hubbard dimer to a Heisenberg model with AF interaction~\cite{auerbach,fazekas99}:
\begin{equation}
  \label{eq:Jkin}
  J_{\rm kin}= \frac{4\,\tau^2}{\cal{U}}
\end{equation}
We thus see that the intermolecular hybridization promotes an effective AF exchange. Of course, if
intermolecular hybridization $\tau$ is comparable to the effective Hubbard ${\cal U}$ the open-shell picture breaks
down and, in the limit where $\frac{{\cal U}}{\tau}$ goes to zero, the dimer becomes a closed shell system. In the
rest of the manuscript we consider the opposite situation.

\subsection{Coulomb-driven superexchange}
\label{sub:cdse}

The second superexchange mechanism involves molecular orbitals different from the ZMs and can be operative even in the absence of
intermolecular hybridization ($\tau=0$). This is particularly relevant in the case of system III in Fig.~\ref{fig:structures}, for
which the two ZMs remain degenerate, on account of the sublattice imbalance of that molecule. 

\subsubsection{Perturbation theory}
\label{subsub:cdse_pt}

The key physical process that drives Coulomb-driven superexchange (CDSE) is encoded 
by the following matrix elements of the Hubbard interaction in the MO basis (\ref{eq:Umat}): 
\begin{equation}
  \mathcal{U}_{+\zeta\zeta-} = U \sum_{i} \psi^\ast_+(i) \, |z_\zeta(i)|^2 \, \psi_-(i) =\mathcal{U}_{-\zeta\zeta+} 
  \label{ME}
\end{equation}
where $\zeta=1,2$ denotes the  ZMs of the building blocks, and $\psi_+$ ($\psi_-$) denotes the LUMO+1 (HOMO-1) orbital.

\begin{figure}
  \includegraphics[width=\linewidth]{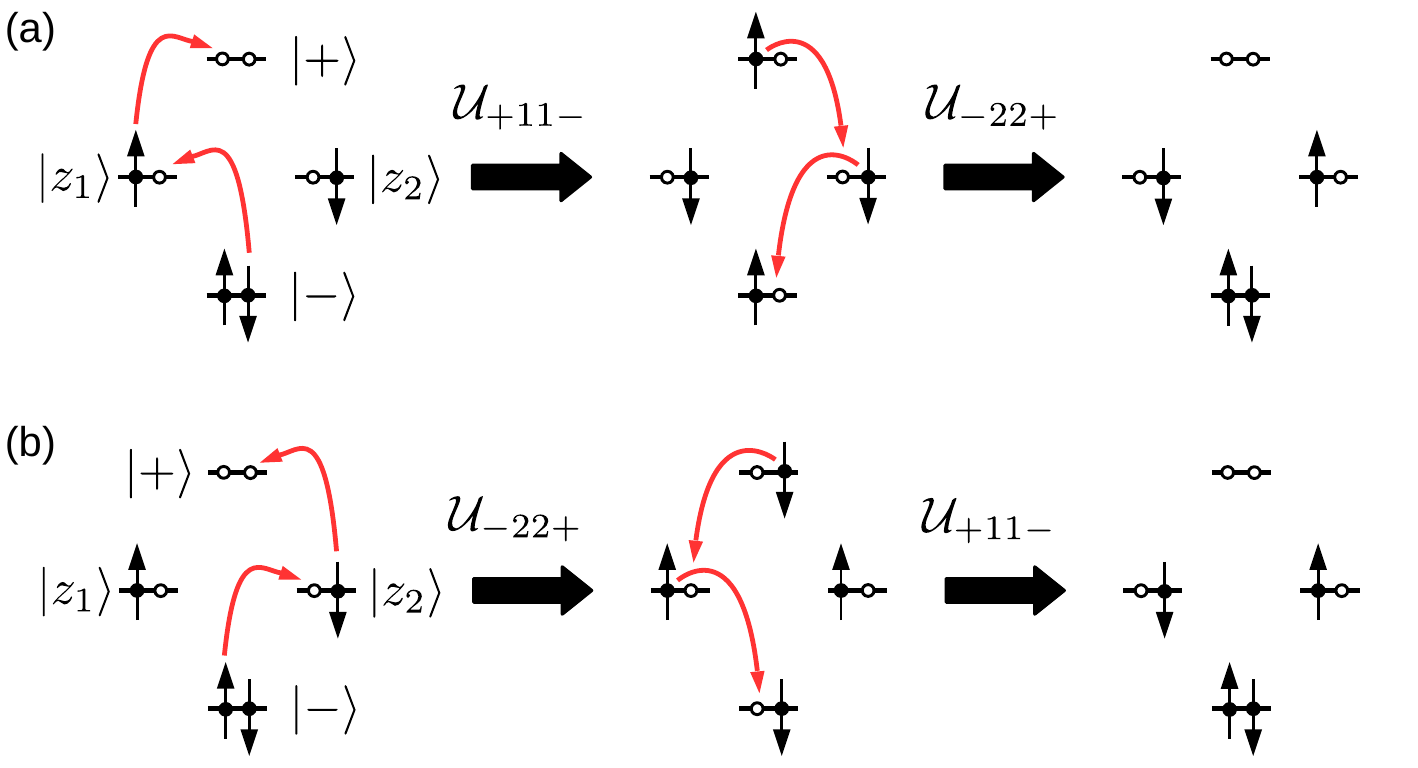}
  \caption{ \label{fig:cdse}
    Coulomb-driven superexchange mechanism mediated by Coulomb matrix elements $\mathcal{U}_{+\zeta\zeta-}=\mathcal{U}_{-\zeta\zeta+}$
    of the Hubbard interaction $\mathcal{H}_U$ given by (\ref{ME}) whereby the spins of the two ZMs are flipped.
  }
\end{figure}

In order to derive how these matrix elements promote exchange we break down the many-body Hamiltonian, represented in the MO basis,
in two parts:
\begin{eqnarray}
  \mathcal{H} &=& \mathcal{H}_0 + \mathcal{H}_1 \\
  \mathcal{H}_0 &=& \sum_{k\in{\rm C}} \epsilon_k \, \hat{N}_{k}
                    + \frac{1}{2} \sum_{k,k^\prime,\sigma,\sigma^\prime} \mathcal{U}_{kk^\prime kk^\prime} \, \hat{N}_{k\sigma} \, \hat{N}_{k^\prime\sigma^\prime}
  \\
  \mathcal{H}_1 &=& \sum_{{k,k^\p,q, \sigma,\sigma^\p} \atop {k\neq k^\p\neq q,\sigma=-\sigma^\p}} \mathcal{U}_{kqqk^\p} \, C_{k\sigma}^\dagger \, C_{q\sigma^\prime}^\dagger C_{k^\p\sigma^\p} C_{q\sigma}
\end{eqnarray}
where in the last line we have taken into account that $\mathcal{U}_{kqqk^\p}=\mathcal{U}_{qkk^\p q}$, thus eliminating the prefactor of 1/2.
The ${\cal H}_0$ term includes both the single-particle energy and the interaction terms that commute with the occupation of the MO.
The second term, ${\cal H}_1$, contains the Coulomb matrix elements that change the occupation of the MOs, giving rise to the excitation of
electron-hole pairs in the non zero modes (see Fig.~\ref{fig:cdse}). We now treat ${\cal H}_1$ as the perturbation. 

The eigenstates of the unperturbed Hamiltonian $\mathcal{H}_0$ can be simply written as Slater determinants,
$\sket{\Phi_m} =C_{k\sigma}^\dagger C_{k^\p\sigma^\p}^\dagger C_{k^\pp\sigma^\pp}^\dagger C_{k^\pp\sigma^\pp}^\dagger \sket{0}$ with energy $E_m^{(0)}$.

For a dimer with two un-hybridized ZMs, the GS of the unperturbed Hamiltonian $\mathcal{H}_0$ is four-fold degenerate.
This is the case of structures I, II, III provided that we take $t_3=0$. 
In the product basis the GS quartet is given by 
\begin{equation}
  \sket{\Phi^{\sigma,\sigma^\p}_0} \equiv C_{1\sigma}^\dagger C_{2\sigma^\p}^\dagger C_{-\up}^\dagger C_{-\dn}^\dagger \sket{0} ,
\end{equation}
i.e., the HOMO-1 is completely filled, and each of the ZMs is carrying one electron of arbitrary spin, while the LUMO+1 is completely empty.
Two of these are the initial and final states of the second order process shown in Fig.~\ref{fig:cdse}. 
For our purposes it is convenient to choose for the GS quartet the eigenstates $\sket{\Psi_{S;S_z}}$
of the total spin $S$ and one of its components $S_z$. First, the $S=0$ state is a singlet: 
\begin{equation}
  \sket{ \Psi_{0;0} } =\frac{1}{\sqrt{2}} [\sket{\Phi^{\up,\dn}_0}-\sket{\Phi^{\dn,\up}_0}]
\end{equation}
On the other hand, the $S=1$ states form a triplet:
\begin{eqnarray}
  \label{eq:ST}
  \sket{ \Psi_{1;-1} } &=& \sket{\Phi^{\dn,\dn}_0} \nonumber \\
  \sket{ \Psi_{1;+0} } &=&\frac{1}{\sqrt{2}} \left[ \sket{\Phi^{\up,\dn}_0} + \sket{\Phi^{\dn,\up}_0} \right] \nonumber \\
  \sket{ \Psi_{1;+1} } &=&\sket{ \Phi^{\up,\up}_0 }
\end{eqnarray}.

If we  consider unhybridized ZMs,   the first order perturbation $\mathcal{H}_1$ does not affect the eigenstates, and thus does not lift the GS degeneracy.
In the case where this assumption fails, as in the case of structure III with finite $t_3$ or when long-range Coulomb interactions are included, the following
results would be modified in two simple ways. First, a splitting of the singlet-triplet manifold linear in the Coulomb repulsion takes place. Second, the denominators in
the second-order perturbation theory  have to be modified accordingly.

Because of the spin-rotational invariance of the many-body Hamiltonian, the representation of the L\"owdin second order degenerate perturbation matrix in the basis of
total spin eigenstates $\ket{\Psi_{S;S_z}}$ is diagonal. As a result, in second order the degeneracy of the spin-singlet and spin-triplet states is lifted according to
\begin{eqnarray}
  J_{\rm cd} &=& E_{S=1}^{(2)} - E_{S=0}^{(2)}=\nonumber\\
             &=& \sum_{m\neq0} \left\{ 
                 \frac{ \left| \sbra{\Phi_m} \mathcal{H}_1 \sket{\Psi_{1;0}} \right|^2 }{E_0^{(0)} - E_m^{(0)} }
                 - \frac{ \left| \sbra{\Phi_m} \mathcal{H}_1 \sket{\Psi_{0;0}} \right|^2 }{E_0^{(0)} - E_m^{(0)} }
                 \right\}
                 \nonumber\\
\end{eqnarray}
where we are comparing the energies of the $S=0$ and the $S=1,S_z=0$ states.
Now, using Eq.~\ref{eq:ST} we can write the Coulomb-driven superexchange splitting in second order as:
\begin{eqnarray}
  \label{eq:JCDper}
  J_{\rm cd}=2 \sum_{m\neq0}  \frac{ \sbra{\Phi^{\dn,\up}_0} \mathcal{H}_1 \sket{ \Phi_m } \sbra{ \Phi_m } \mathcal{H}_1  \sket{\Phi^{\up,\dn}_0} }{ E_{\rm gs}^{(0)} - E_m^{(0)} }
\end{eqnarray}
Hence, as illustrated in Fig.~\ref{fig:cdse},
the energy difference between singlet and triplet states comes from the effective spin-flip
$\sket{\Phi^{\dn,\up}_0}\longrightarrow \sket{\Phi^{\up,\dn}_0}$
via intermediate (virtual) states where the LUMO+1 becomes occupied,
i.e., $\sket{\Phi_m}=C^\dagger_{+\sigma} C^\dagger_{\zeta\bar\sigma} C_{-\bar\sigma} C_{\zeta\sigma} \sket{\Phi_0^{\sigma,\bar\sigma}}$,
where $\zeta=1,2$.

Plugging in the matrix elements $\sbra{\Phi_m^{\vspace{2ex}}}\mathcal{H}_1\sket{\Phi^{\dn,\up}_0}\sim\mathcal{U}_{+\zeta\zeta-}$
and taking into account that there are only two second order processes connecting $\sket{\Phi_0^{\up,\dn}}$ and $\sket{\Phi_0^{\dn,\up}}$
(the ones shown in Fig.~\ref{fig:cdse}), we finally obtain for the Coulomb-driven superexchange in second order:
\begin{equation}
  \label{eq:JCD}
  J_{\rm cd} = 4\cdot\frac{\mathcal{U}_{+11-}\cdot\mathcal{U}_{-22+}}{-\Delta_{+-}}
\end{equation}
where $\Delta_{+-}\equiv\epsilon_+-\epsilon_-$ is the energy gap between LUMO+1 and HOMO-1, which is of the order of $t$.

\subsubsection{Sign of Coulomb-driven kinetic exchange}
\label{subsub:cdse_sign}

We now discuss how Eq.~(\ref{eq:JCD}) can lead both to FM and AF superexchange.
The key observation is the following identities. In the case of systems I and II, we find:
\begin{equation}
  \label{symAF}
  \mathcal{U}_{+11-}=-\mathcal{U}_{+22-}
\end{equation}
In  contrast, in the case of system III we find:
\begin{equation}
  \label{symFM}
  \mathcal{U}_{+11-}=+\mathcal{U}_{+22-}
\end{equation}

These results can be understood in terms of the relation of the molecular orbitals in
of electron-hole symmetric bipartite systems~\cite{ortiz19}:
\begin{equation}
  \psi_{\pm}=\left(\begin{array}{c} \psi_A \\ \pm \psi_B \end{array}\right)
\end{equation}
that shows that the MOs $\psi_+$ and $\psi_-$ are the bonding and anti-bonding states
of two identical sublattice polarized orbitals.

For the AF case,  the 
ZMs $1,2$ belong to different sublattices, $A$ and $B$, respectively, and we may write:
\begin{equation}
  \mathcal{U}_{+11-}=U\sum_i \psi_A(i)^* |z_1(i)|^2 \psi_A(i)
\end{equation}
and
\begin{equation}
  \mathcal{U}_{+22-}= U \sum_i \psi_B(i)^* |z_2(i)|^2 (-1) \psi_B(i)
\end{equation}
Now we use the fact that the inversion symmetry of molecules I and II is identical to sublattice inversion and we obtain Eq.~(\ref{symAF}).

For the FM case, both ZMs $z_1$ and $z_2$  belong to the same sublattice, that we decide to label as $A$. 
We thus write the matrix elements in Eq.~(\ref{symFM}) as
\begin{equation}
  \mathcal{U}_{+11-}=U\sum_i \psi_A(i)^* |z_1(i)|^2 \psi_A(i)
\end{equation}
and
\begin{equation}
  \mathcal{U}_{+22-}=U\sum_i \psi_A(i)^* |z_2(i)|^2 \psi_A(i)
\end{equation}
Now we use the fact that orbitals $z_1$ and $z_2$ are related by a mirror symmetry so that these two matrix elements have to be identical. 

The matrix elements (ME) describe non-diagonal exchange processes, where an electron hops from
one of the ZMs to the LUMO+1 and an electron of opposite spin coming from the HOMO-1 hops to the emptied ZM,
as well as the inverse process.
The combination of two such \emph{exchange hoppings} gives rise to the second-order process that leads to an
effective spin flip between the two ZMs, as illustrated in Fig.~\ref{fig:cdse}. Ultimately, the sign
is governed by the presence of a relative sign in the sublattice .

\section{Numerical results}
\label{sec:numeric}

In this section we support the theoretical framework presented so far with  numerical results for the multi-configurational calculations
of the three dimers of Fig.~\ref{fig:structures} using the Hubbard model. The effect of LR Coulomb interaction will be discussed in Sec.~\ref{sec:lr_coulomb}.
We make no attempt to derive the value of  $U$ from first principles~\cite{wehling11} and we choose instead to plot excitation energies as a function of $U$. We
note that $U$ will depend not only on the molecule but also on the substrate, due to screening of Coulomb repulsion. Therefore, $U$ can be tuned,
to some extent.

\subsection{Choice of active space to isolate superexchange mechanisms}
\label{sub:cas_choice}

Importantly, the numerical method permits us to switch on and off the different intermolecular exchange mechanisms.
If we restrict the active space to the minimal space of non-bonding ZMs, we automatically disable the Coulomb-driven IE.
By the contrary, in the case of disconnected ZMs, where intermolecular hybridization is only allowed by third-neighbor
hopping, we can take $t_3=0$, to disable the kinetic exchange  and include non-zero modes in the active space to allow for
Coulomb-driven IE. We can thus study the two different superexchange mechanisms independently.

\subsection{AF phenalenyl dimers}
\label{sub:af_dimers}

We now consider the two phenalenyl-diradicals I and II shown in Fig.~\ref{fig:structures}. Both are   dimers
made of two small $S=1/2$  triangulenes. The lattice imbalance of these  phenalenyl dimers is $|N_A-N_B|=0$. Thus the GS
is expected to be a spin singlet ($S=0$), anticipating an AF coupling between the two spin-1/2 phenalenyl units. 
The main difference between these two structures should be related to the smaller intermolecular hybridization of
structure II, reflected in the single-particle spectrum [see Eq.~(\ref{eq:deltasp})].

\begin{figure}
  \includegraphics[width=\linewidth]{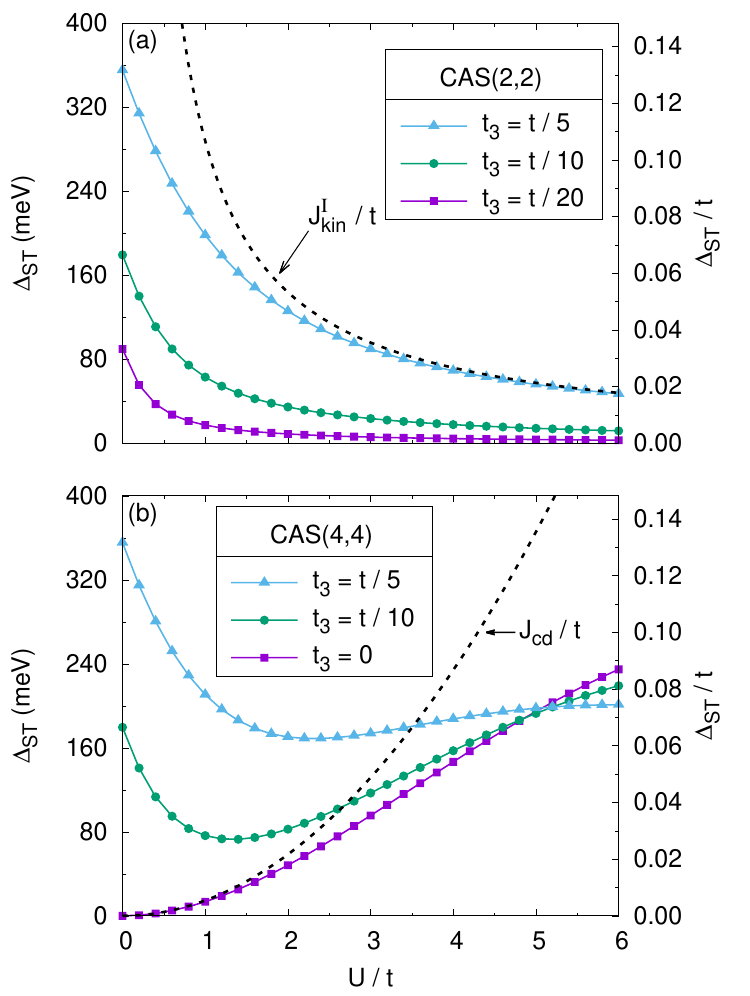}
  \caption{
    \label{fig:minidimer_exch}
    Singlet-triplet splitting $\Delta_{\rm ST}$  for the Hubbard model
    of phenalenyl dimer I shown in Fig.~\ref{fig:structures} 
    in the absence of LR Coulomb interactions ($V_{i,j}=K_{i,j}=0$)
    as a function of Hubbard $U$ computed for (a) CAS(2,2) and
    (b) CAS(4,4) for different values of $t_3$.
    The black dashed line in (a) shows $J^{\rm I}_{\rm kin}$
    according to (\ref{eq:JkinI}) for $t_3=t/5$. The black dashed
    line in panel (b) shows $J_{\rm cd}$ given by (\ref{eq:JCD}).
     The right vertical axis shows $\Delta_{\rm ST}$ in units of $t$, while the left axis
      shows $\Delta_{\rm ST}$ in meV assuming $t=2.7$eV.   
  }
\end{figure}

We start the discussion with CAS(2,2), the minimal Hilbert space, that includes only two  electrons in two ZMs.
We obtain our results numerically and compare with the analytical results obtained in Sec.~\ref{sub:kin_exch}.

For all values of $U$, the ground state is a singlet and the first excited state is a triplet.  In the following we focus on their energy difference:
\begin{equation}
  \label{eq:deltast}
  \Delta_{\rm ST}\equiv E_{S=1}-E_{S=0}
\end{equation}
as a function of $U/t$. This quantity is important because it can be measured using inelastic electron tunneling spectroscopy
(IETS)~\cite{li2020,mishra2020,ortiz20,mishra21a,mishra21b,hieulle2021,cheng22,de2022}. In Fig.~\ref{fig:minidimer_exch}(a) we show
$\Delta_{\rm ST}$ for the direct phenalenyl dimer (structure I) for three different values of $t_3$. From comparison of tight-binding
and first-principles calculations~\cite{ortiz22} for triangulene crystals we believe $t_3\simeq 0.1 t$ is a good estimate. The
values of $U/t$ that provide good agreement between the predictions of the Hubbard model for NGs and IETS experiments
are in the range of $1.5<U/t<2$ but this is to be taken as a rough estimate.

For $U=0$, the singlet-triplet splitting equals the single-particle splitting of the ZMs, shown in Fig.~\ref{fig:deltasp},
since the only way to have two parallel spins is to have them in different orbitals. As we ramp up $U$ the wave function of the
ground state evolves adiabatically from a closed-shell singlet, in which states of doubly occupied ZMs have the same
weight as those with single occupancy, to an open-shell singlet, where the weight of doubly occupied ZMs is
significantly reduced~\cite{ortiz19}. The figure of merit that controls this crossover is $r=\mathcal{U}/\tau$,
defined in Eqs.~(\ref{eq:tau}) and (\ref{eq:Ueff}). In the strong coupling limit ($r\gg1$), the singlet-triplet splitting
is given by Eq.~(\ref{eq:Jkin}).

The results for  structure II (not shown) are qualitatively identical to those of structure I, but with a smaller value of the
singlet-triplet splitting for all values of $U$ and $t_3$, on account of the smaller intermolecular hybridization shown in
Fig.~\ref{fig:deltasp}. The ZM splitting (\ref{eq:deltasp}) relates to the value of the effective $\tau$ via $\delta=2\tau$.
Thus, for $U=0$ the splittings are given by $2t_3/3$ for structure I and by $\simeq0.72t_3^2/t$ for structure II.
The effective Hubbard-U, $\mathcal{U}$, remains almost independent of $t_3$ for structures I and II with $\mathcal{U}\simeq\frac{1}{6}$
for the relevant values of $t_3$

In the strong coupling limit, the kinetic exchange for structures I and II is given by:
\begin{eqnarray}
  \label{eq:JkinI}
  J_{\rm kin}^{\rm I} &=& \frac{8}{3}\,\frac{{t_3}^2}{U} \\
  \label{eq:JkinII}
  J_{\rm kin}^{\rm II}& =& \left(\frac{0.71 t_3}{t}\right)^2\frac{6{t_3}^2}{U} =
                           \left(\frac{0.71 t_3}{t}\right)^2 \frac{9}{4} J_{\rm kin}^{\rm I}
\end{eqnarray}

We now study Coulomb-driven superexchange by enlarging the Hilbert space in order to  take 
into account the HOMO-1 and LUMO+1 in addition to the ZMs.  This defines an active space of four electrons in four states 
[CAS(4,4) approximation]. This is the minimal space in which Coulomb-driven superexchange  can appear. The large
degeneracy of both LUMO+2 and HOMO-2 manifolds moves the dimension of the Hilbert space out of our computational capability.
However, we can validate the analytical perturbation formulas with CAS(4,4) and then use perturbation theory to include
higher energy MOs (see Sec.~\ref{sec:full_pt}).
 
For CAS(4,4) the GS also has $S=0$ and the first excited state $S=1$ for both structures I and II.
In Fig.~\ref{fig:minidimer_exch}(b) we plot $\Delta_{\rm ST}$ as a function of $U$, for three values of $t_3$. For $t_3=0$ the
kinetic exchange is eliminated. As a result, the $t_3=0$  curve shows the magnitude of the CDSE
associated to the virtual processes of Fig.~\ref{fig:cdse}.  It is apparent that, for moderate values of $U/t<2$,
the exact calculation and the perturbation theory result [dashed line in\ref{fig:minidimer_exch}(b)] are in good agreement.
This validates the perturbation theory.

The excitation energies with finite $t_3$ and CAS(4,4) are very well approximated by the sum of the CAS(2,2) and
the $t_3=0$ curve of CAS(4,4), as long as $t_3$ remains much smaller than $t$.  Therefore, we conclude that these contributions
are additive, for moderate values of $U/t$. Since kinetic exchange and CDSE have opposite dependence with $U$,
the resulting  $\Delta_{ST}$  curves are non-monotonic functions of $U$. For small values of $U$ they are dominated by the single-particle
splitting. As $U$ increases, local moments develop, CDSE takes over, and kinetic exchange fades away according
to Eq.~(\ref{eq:Jkin}).

\begin{figure}
  \includegraphics[width=\linewidth]{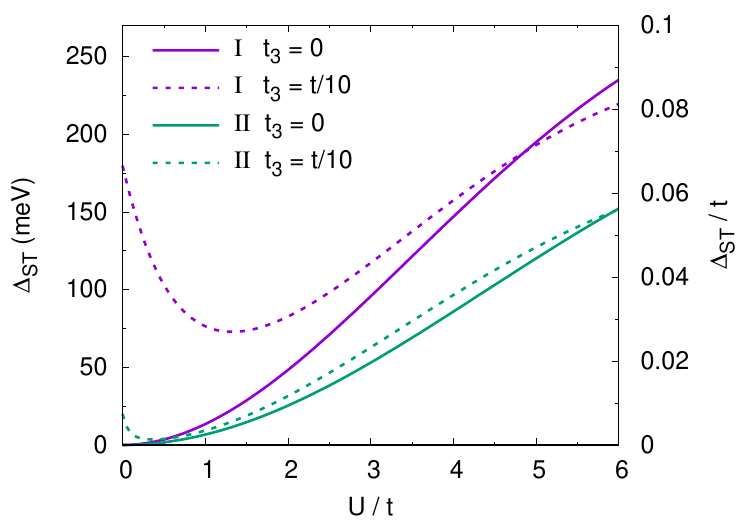}
  \caption{ \label{fig:comp12}
    Comparison of singlet-triplet splittings $\Delta_{\rm ST}$ for Hubbard models of phenalenyl
    dimers I and II computed for CAS(4,4) for different values of $t_3$ as a function of $U$.
    The right vertical axis shows $\Delta_{\rm ST}$ in units of $t$, while the left axis
    shows $\Delta_{\rm ST}$ in meV assuming $t=2.7$eV.}
\end{figure}

We now compare the excitation energies $\Delta_{\rm ST}$ for structures I and II in the CAS(4,4) approximation for different values of $t_3$. The results
are shown in Fig.~\ref{fig:comp12}. It is apparent that, for all values of $U$  $\Delta_{\rm ST}$ is smaller
for structure II than for structure I, for a given value of $t_3$. This can be understood as follows.
First, kinetic exchange is smaller for structure II on account of the smaller intermolecular hybridization $\tau$, see Fig.~\ref{fig:deltasp}.
As a result, the crossover to the open-shell regime
happens for much smaller values of $U/t$ and the kinetic exchange ($\propto\tau^2/\mathcal{U}$) is dramatically smaller.
Second, also the CDSE is considerably smaller for structure II than for structure I, as is evident by comparing the results of both structures
for $t_3=0$ in Fig.~\ref{fig:comp12} when the kinetic superexchange is absent.
The reason is the smaller overlap of the ZMs with the HOMO-1 and LUMO+1 orbitals in structure II than in structure I
[cf. Figs.~\ref{fig:singleparticle}(d,e)], leading to smaller Coulomb matrix elements $\mathcal{U}_{+\zeta\zeta-}$ and consequently
smaller CDSE.

\subsection{FM phenalenyl dimer}
\label{sub:fm_dimer}

We now discuss the case of structure III where, as in structure II, two phenalenyl molecules
are coupled indirectly via the central benzene molecule. But in contrast to structure II
the two phenalenyl units now form a 120 degree angle instead of 180 degrees; see Fig.~\ref{fig:structures}.
In this case the lattice imbalance of the combined system is $|N_A-N_B|=2$. Thus by Lieb's
theorem the GS is expected to have spin $S=1$, anticipating a \emph{FM} IE between
the spin-1/2 of individual phenalenyl units. The crucial difference with AF phenalenyl
dimers considered before in Sec.~\ref{sub:af_dimers} is that now the two ZMs localized
at the edges of the phenalenyl units are on the same sublattice. 

Figure~\ref{fig:singleparticle}(c) shows the single-particle spectrum
obtained from diagonalization of the hopping Hamiltonian (\ref{eq:Ht}) in the absence of third
neighbor hopping ($t_3=0$). As in the case of structures I and II there are two ZMs, each
localized on one of the phenalenyl units, shown in the two center panels of Fig.~\ref{fig:singleparticle}(e).
In contrast to structures I and II, the HOMO-2, LUMO+2 are non-degenerate, so that we can include
them in the  active space and check the convergence of the energy spectrum as more MOs are included. Their wave
functions are shown in Fig.~\ref{fig:singleparticle}(f).

\begin{figure}
  \includegraphics[width=\linewidth]{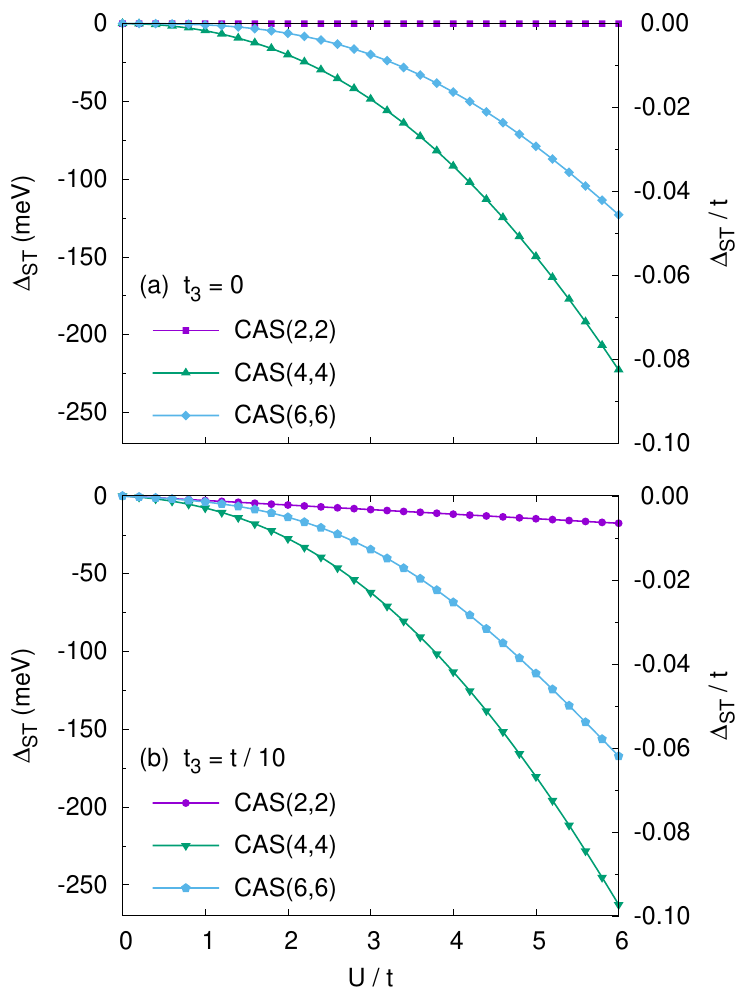}
  \caption{
    \label{fig:fmdimer_exch}
    Singlet-triplet splitting $\Delta_{\rm ST}$  for Hubbard model of 
    phenalenyl dimer III shown in Fig.~\ref{fig:structures}
    as a function of $U$ computed for CAS(2,2), CAS(4,4) and CAS(6,6), and for
    (a) $t_3=0$ (to panel), and (b) $t_3=0.1t$ (bottom panel). The right vertical axis shows $\Delta_{\rm ST}$
    in units of $t$, while the left axis shows $\Delta_{\rm ST}$ in meV assuming $t=2.7$eV.
  }
\end{figure}

In the CAS(2,2) approximation, taking into account only the two ZMs, only the Hund's exchange is active,
as the intermolecular hybridization is warranted to vanish on account of the sublattice imbalance. For $t_3=0$
there is no overlap of the ZMs so that $J_{\rm Hund}=0$.
Turning $t_3$ on keeps the ZMs degenerate, but their wave functions spread a bit and overlap, giving rise
to finite Hund's exchange (see purple curves in Fig.~\ref{fig:fmdimer_exch}). In the top panel,
with $t_3=0$, $\Delta_{\rm ST}$ is strictly zero. On the other hand, for finite $t_3$ shown in the bottom panel, we
find a finite $\Delta_{\rm ST}$ approximately linear in $U$.
The negative sign accounts for the fact that the GS has $S=1$ and the first excited state has $S=0$,
complying with Lieb theorem. Thus structure III has \emph{FM} intermolecular exchange. We note
that $J_{\rm Hund}=0$ for structures I and II in the Hubbard approximation, as there the ZMs live in opposite
sublattices and thus the ZMs have zero overlap.

Taking into account the HOMO-1 and LUMO+1 in addition to the ZMs leads to a CAS(4,4) calculation
and enables CDSE. We find that, even for $t_3=0$, a spin excitation gap opens, as can be seen in
Fig.~\ref{fig:fmdimer_exch}(a). It is apparent that the CDSE mediated by the HOMO-1 and  LUMO+1
intermediate states gives a \emph{FM} contribution, different from structures I and II discussed
in the previous subsection. For CAS(4,4) and finite $t_3$ the contributions coming from the direct exchange
mechanism linear in $U$ and from the FM superexchange quadratic in $U$ approximately add up, giving
rise to approximately linear plus quadratic behavior, as can be seen in Fig.~\ref{fig:fmdimer_exch}(b).

On the other hand, including also the HOMO-2 and LUMO+2 MOs in a CAS(6,6) calculation, we find that the spin
excitation gap decreases with respect to CAS(4,4); see the green and blue curves in Figs.~\ref{fig:fmdimer_exch}(a,b).
Thus, the contribution of these additional MOs to the CDSE is of opposite sign than the one of the HOMO-1 LUMO+1 pair,
i.e., it is AF, but smaller in magnitude, so that overall, the CDSE remains FM.
This sign can be understood again by looking at the phases of all intermediate MOs in Fig.~\ref{fig:singleparticle}(f).
Now the phases of the combined wave functions $\psi_{+}(i)\,\psi_{-2}(i)$ and $\psi_{-}(i)\,\psi_{+2}(i)$ have different signs
in both phenalenyl units. Since the energy difference between HOMO-2 and LUMO+1, and HOMO-1 and LUMO+2, is larger
than the energy difference between HOMO-1 and LUMO+1, $\epsilon_{+}-\epsilon_{-2}=\epsilon_{+2}-\epsilon_{-}>\Delta{+-}$,
this AF contribution to the exchange coupling is smaller in magnitude
than the FM contribution.

Finally, we note  that the two FM exchange mechanisms described here probably 
account for the formation of the high-spin GS in a triangulene-trimer coupled via a central
benzene ring, reported recently in Ref.~\onlinecite{cheng22}.

\section{Perturbation theory of superexchange with entire single-particle spectrum}
\label{sec:full_pt}

The comparison of CAS(4,4) and CAS(6,6) for structure III shows a large variation of the singlet-triplet splitting $\Delta_{\rm ST}$.
The inclusion of additional single-particle states in the CAS calculation faces the problem of exponential growth of the Hilbert space.
On the other hand, the comparison of results for CAS(4,4) with perturbation theory for structure I shows very good agreement between these
two methods for $U<1.5t$, when only the HOMO-1, LUMO+1 orbitals are included in the perturbative calculation.
Therefore, we now calculate the perturbative result including {\em all MOs} of the single-particle spectrum.
This allows us to assess the contributions of excited states that are excluded from the active space. 

The generalization of Eq.~(\ref{eq:JCD}) for an arbitrary number of excited states is straight-forward:
\begin{equation}
  \label{eq:JCD_full}
  J_{\rm cd}^{\rm full} = -4 \sum_{{k_+=+1,+2,\ldots} \atop {k_-=-1,-2,\ldots}} \frac{\mathcal{U}_{k_+11k_-}\mathcal{U}_{k_-22k_+}}{\epsilon_{k_+}-\epsilon_{k_-}}
\end{equation}
Note that the number of terms in this sum only grows quadratically with the number of excited single-particle states,
compared to the exponential scaling of CAS(N,N) calculation.

\begin{figure}
  \includegraphics[width=\linewidth]{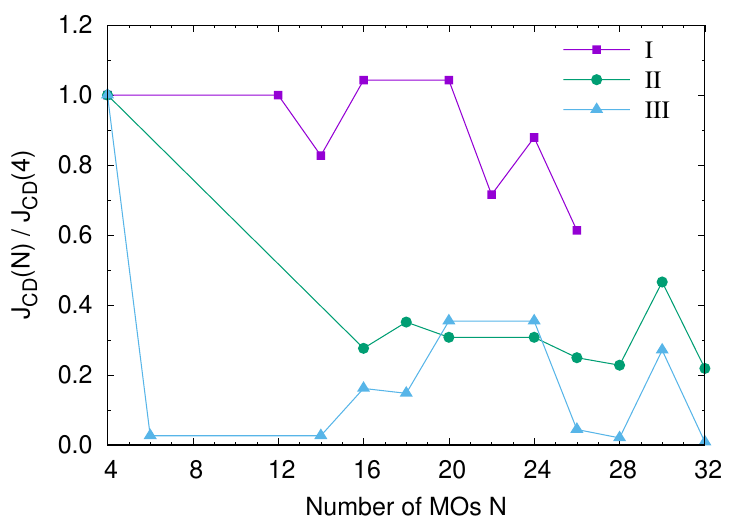}
  \caption{\label{fig:JCD_PT}
    Coulomb-driven superexchange coupling $J_{\rm cd}$ computed in second order perturbation theory
    according to Eq.~(\ref{eq:JCD_full}) for all three structures (I, II, III in Fig.~\ref{fig:structures})
    as a function of the number $N$ of MOs taken into account in the summation.
  }
\end{figure}

The results for structures I, II, and III are shown in Fig.~\ref{fig:JCD_PT}. We plot $J_{\rm cd}$ as a function of the number of  MO
pairs included in the calculation. Note that these increase at least by two units, on account of the electron-hole symmetry, and in
some cases in larger steps, on account of the degeneracies of the single-particle spectrum associated to the symmetries of the molecules.
We plot $J_{\rm cd}$ normalized to the contribution of the HOMO-1/LUMO+1 pair.

We observe that including more orbitals may increase or decrease the magnitude of
the exchange coupling, indicating that the extra contributions coming from adding new orbitals
may have different signs. This can be understood again by analyzing the symmetries of the
MOs involved in electron-hole pair excitations w.r.t. the sublattice polarization of the ZMs.
We note that terms with ``diagonal'' electron-hole pair excitations $k_-\rightarrow{k}_+=-k_-$
always have the same sign as the lowest order term ($N=4$), i.e., AF for structures I and II,
and FM for structure III.

The change in magnitude of including MOs beyond the HOMO-1/LUMO+1 pair cannot be
neglected. Especially in the FM case, structure III, the exchange coupling changes
substantially right up to the inclusion of the last term. Note that although the magnitude of the
exchange coupling for this structure decreases by two orders of magnitude compared to only
including the HOMO-1/LUMO+1 contribution, it does not become zero and remains FM.
Also for the two AF structures (I,II) the exchange coupling keeps changing right
to the end, although not as drastically as in the FM case (III). In the case of structure
I the magnitude of the exchange coupling is reduced to about 60\%, and in the case of structure II
to about 20\% of its value at $N=4$. 

Hence contributions from electron-hole pair excitations of higher energy MOs to the CDSE can generally
not be neglected, at least on a quantitative level. It should be noted, however, for structure I
in the experimentally relevant parameter regime ($t=2.7$eV, $t_3=t/10$ and $U\sim{t}$), kinetic
superexchange is the dominant mechanism ($J_{\rm kin}\sim70$meV) compared to CDSE ($J_{\rm cd}\sim15$meV).
On the other hand, our generalized perturbation expression (\ref{eq:JCD_full}) allows us to actually
calculate the missing contribution $\delta{J}_{\rm cd}$ to the CDSE coupling due to higher energy orbitals, which could then be
incorporated e.g. by an effective Heisenberg interaction, $\delta{J}_{\rm cd}\,\vec{S}_1\cdot\vec{S}_2$,
into our model, similar to our earlier work~\cite{jacob21}.

\section{The role of LR Coulomb interactions}
\label{sec:lr_coulomb}

So far we have treated Coulomb interaction in the Hubbard approximation. We now address the role
of  LR Coulomb interactions~\cite{shi17} on the spin excitations and consider the extended Hubbard model,
including the LR part of the Coulomb repulsion and direct Coulomb exchange given by (\ref{eq:Coul})
and (\ref{eq:Exch}), respectively. 

In order to obtain realistic parameters for $V_{i,j}$ and $K_{i,j}$ \emph{relative} to the on-site Hubbard
interaction $U$, we have computed the full Coulomb matrix for the $\pi$-orbitals $\phi_i(r)$ of the
carbon atoms using the quantum chemistry code Gaussian09~\cite{G09} and a minimal basis set (see
the Appendix for more details). This yields the bare Coulomb interaction
$v^{\rm bare}_{ijkl}=\bra{\phi_i,\phi_j}\hat{v}_c\ket{\phi_k,\phi_l}$ where $\hat{v}_c=1/r$. 
However, the effective Coulomb interaction for our $\pi$-orbital only model must
generally be considerably lower than the bare one due to screening by the other orbitals.
On the lowest level of approximation we can include screening effects by a dielectric
constant $\epsilon$, scaling all matrix elements equally, $v^{\rm scr}_{ijkl}=\epsilon^{-1}\,v_{ijkl}^{\rm bare}$.
Specifically, in this approximation the on-site Hubbard interaction $U$ is given by
$U=v^{\rm scr}_{iiii}=\epsilon^{-1}\,v_{iiii}^{\rm bare}$ and hence $\epsilon^{-1}=U/v_{iiii}^{\rm bare}$.
We can then parametrize the other interactions in terms of the Hubbard $U$,
namely the intersite repulsion becomes $V_{i,j}=v^{\rm scr}_{ijij}=(U/v^{\rm bare}_{iiii})\,{v_{ijij}^{\rm bare}}$
and the direct intersite exchange becomes $K_{i,j}=v^{\rm scr}_{ijji}=(U/v^{\rm bare}_{iiii})\,{v_{ijji}^{\rm bare}}$.
This allows us to continue using the Hubbard $U$ as a parameter in our model.
But effectively it means we are tuning the screening parameter $\epsilon$,
allowing us to simulate in an approximate way the screening effect of different
substrates on the LR part of the Coulomb interaction and exploring its effect
on the spin excitation gap.

\begin{figure}
  \includegraphics[width=0.95\linewidth]{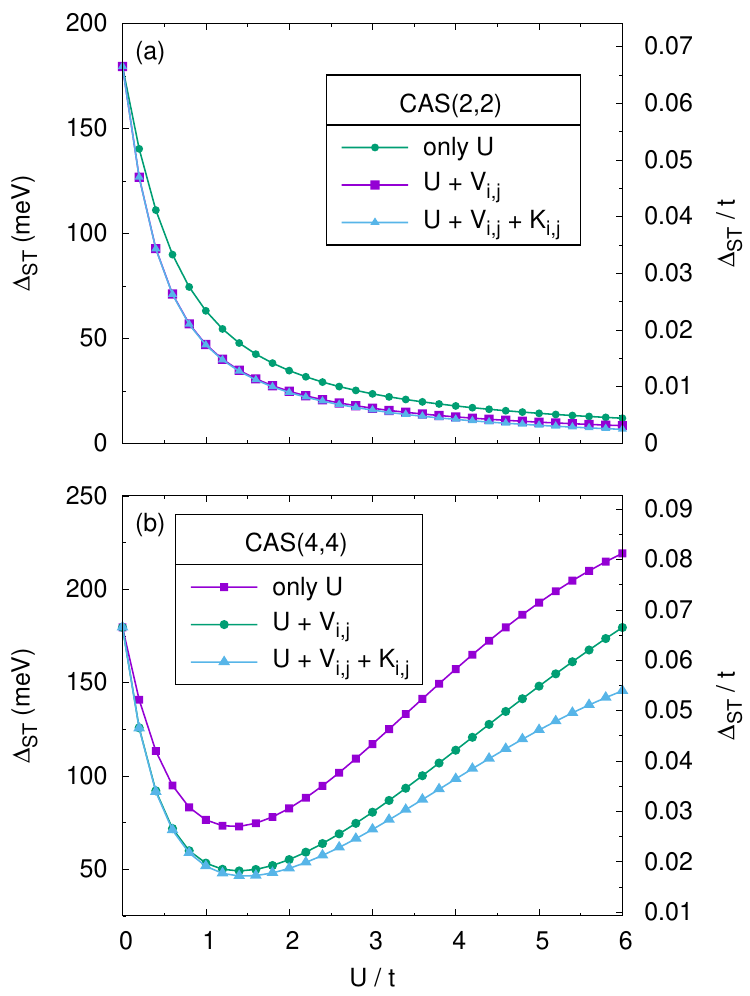}
  \caption{\label{fig:minidimer_lr}
    Comparison of singlet-triplet splitting $\Delta_{\rm ST}$ for the extended Hubbard model of phenalenyl
    dimer (structure I) including different parts of the Coulomb interaction as a 
    function of Hubbard $U$ computed for (a) CAS(2,2) and  (b) CAS(4,4) and for
    $t_3=0.1t$. All interactions are scaled proportional to the Hubbard $U$; see
    main text. The right vertical axis shows $\Delta_{\rm ST}$ in units of $t$,
    while the left axis shows $\Delta_{\rm ST}$ in meV assuming $t=2.7$eV.
  }
\end{figure}

\subsection{Effect of LR Coulomb on kinetic exchange}
\label{sub:lr_kin_exch}

We first discuss the effect of LR Coulomb interaction on the kinetic exchange mechanism.
Figure~\ref{fig:minidimer_lr}(a) shows the spin excitation gap for CAS(2,2) (only kinetic
exchange is active). Clearly, by inclusion of the intersite repulsion
$V_{i,j}$ into the model the kinetic exchange is reduced, while the inclusion of direct
exchange $K_{i,j}$ does not have any effect. The latter is easily understood, since the
direct exchange decays exponentially and thus is only appreciable for nearest neighbors;
see Fig.~\ref{fig:lr_coulomb}. But since the sites making up the ZMs are third neighbors
at least, see Fig.~\ref{fig:singleparticle}(d), direct exchange cannot be effective in this case.

For the Hubbard model, the kinetic exchange is given by Eq.~(\ref{eq:Jkin}). It is straight-forward
to generalize this expression for the extended Hubbard model including intersite Coulomb repulsion $V_{i,j}$.
In the presence of LR repulsion the energy cost for transferring an electron from one molecule to the other
becomes 
\begin{equation}
  \delta\mathcal{E} = \mathcal{U} + \mathcal{V}_{11} - \mathcal{V}_{12} > \mathcal{U}
\end{equation}
where ${\cal V}_{\zeta\zeta}=\sum_{i\ne{j}} V_{i,j} |z_\zeta(i)|^2 |z_\zeta(j)|^2$ is the
contribution of the LR repulsion to the \emph{intra}-orbital Coulomb repulsion (or effective Hubbard-U)
within a localized ZM, while $\mathcal{V}_{12}=\sum_{i,j} V_{i,j} |z_1(i)|^2 |z_2(j)|^2$ is the
\emph{inter}-orbital Coulomb repulsion between two ZMs, localized on different molecules. 
 
Thus the LR repulsion leads to an increase of the double occupancy energy compared to
simply $\mathcal{U}$ in the Hubbard model, since $\mathcal{V}_{\zeta\zeta}>\mathcal{V}_{12}$ always, as sites $i,j$ on the different
molecules are farther away \emph{on average} than when both sites are on the same molecule.
In the presence of LR Coulomb repulsion the kinetic exchange (\ref{eq:Jkin}) is thus reduced in comparison to the
Hubbard model according to
\begin{equation}
  J^{\rm LR}_{\rm kin} = \frac{4\tau^2}{\mathcal{U}+\mathcal{V}_{\zeta\zeta}-\mathcal{V}_{12}} < J_{\rm kin} = \frac{4\tau^2}{\mathcal{U}}  
\end{equation}

\subsection{Effect of LR Coulomb on Coulomb-driven superexchange}
\label{sub:lr_cdse}

We now discuss the effect of LR Coulomb interactions on the CDSE.
Figure~\ref{fig:minidimer_lr}(b) compares the spin excitation gap for CAS(4,4) as a function of
the Hubbard interaction $U$ when different parts of the Coulomb interaction are included.
Including the intersite repulsion $V_{i,j}$ reduces the spin excitation gap considerably,
similar to the case of CAS(2,2) considered before when only kinetic exchange is active,
but even more strongly so. Different from the case of CAS(2,2), now also the inclusion of
direct exchange $K_{i,j}$ reduces the spin excitation gap. 

What is the reason behind the reduction of the CDSE by the LR part of the Coulomb repulsion $V_{i,j}$
and the direct exchange $K_{i,j}$?  First, since the densities of the HOMO-1 and LUMO+1 orbitals are the same,
i.e., $|\psi_+(i)|^2=|\psi_-(i)|^2$, it is easy to see that the energy difference between the GS (HOMO-1 full
and LUMO+1 empty) and the intermediate excited states (HOMO-1 and LUMO+1) half-filled, entering in the
denominator in (\ref{eq:JCD}) is not affected by the LR Coulomb repulsion $V_{i,j}$.
On the other hand, direct exchange $K_{i,j}$ increases the energy of the intermediate state, since the spins in the
HOMO-1 and LUMO+1 are antiparallel to the spins of both ZMs, see Fig.~\ref{fig:cdse}, leading to
an energy penalty of 
$4\,\mathcal{K}_{\zeta++\zeta}=4\,\mathcal{K}_{\zeta--\zeta}$ (where $\zeta=1,2$ denotes the ZM) 
due to the FM nature of the direct exchange. Hence the denominator in (\ref{eq:JCD}) is altered according
to $E_0^{(0)}-E_m^{(0)}=-\Delta_{+-}-4\mathcal{K}_{\zeta++\zeta}$,
leading to a reduction of $J_{\rm cd}$.

On the other hand, the "exchange hopping"  matrix elements of the Coulomb interaction
driving the superexchange are obviously directly affected by the introduction of LR repulsion:
$\mathcal{W}_{+\zeta\zeta-}=\mathcal{U}_{+\zeta\zeta-}+\mathcal{V}_{+\zeta\zeta-}$ where $\zeta=1,2$
denotes ZM1 or ZM2, respectively.
Importantly, the LR corrections $\mathcal{V}_{+\zeta\zeta-}=\sum_{i<j}V_{i,j}\,\psi^\ast_+(i)\psi_\zeta^\ast(j)\psi_\zeta(i)\psi_-(j)$
have just the opposite sign of the corresponding matrix elements of the Hubbard interaction
$\mathcal{U}_{+\zeta\zeta-}=U\sum_i \psi^\ast_+(i) |\psi_\zeta(i)|^2\psi_-(i)$
such that $|\mathcal{W}_{+\zeta\zeta-}|<|\mathcal{U}_{+\zeta\zeta-}|$.
The reason behind the opposite signs of Hubbard and LR contributions to the 
"exchange hopping" 
matrix elements are the alternating phases of the ZM wave functions, see Fig.~\ref{fig:singleparticle}d.
Hence the CDSE is also reduced by the LR part of the
Coulomb repulsion compared to (\ref{eq:JCD}).
In summary, CDSE is altered by both LR Coulomb repulsion and direct exchange:
\begin{eqnarray}
  \label{eq:JCD_LR}
  J_{\rm cd}^{\rm LR} &=& 
  4\,\frac{|\mathcal{U}_{+11-}+\mathcal{V}_{+11-}|^2}{\Delta_{+-}+4\,{\mathcal{K}_{1++1}}} < J_{\rm cd}
\end{eqnarray}
where we have already taken into account that 
$\mathcal{W}_{+11-}=-\mathcal{W}_{+22-}$.

Finally, we note that the influence of the LR part of the Coulomb interaction on the spin excitation energies
has a similar effect for structures II and III for CAS(2,2) and CAS(4,4) (not shown). Even though the LR Coulomb
interaction between both molecules is weaker in these cases, since both molecules are farther away, the \emph{relative}
change is similar to structure I, as both kinetic exchange and CDSE are also reduced by the larger distance
between the coupled molecules.

\section{Discussion and Conclusions}
\label{sec:conclusions}

The main goal of this work is to provide a framework to understand intermolecular exchange interactions for $S=1/2$ triangulenes. However,  the results of this paper can be directly applied to other $S=1/2$ NGs, such as Clar's goblet~\cite{mishra2020c}. We have discussed three different intermolecular exchange interactions. We have shown how intermolecular hybridization and the sub-lattice imbalance determine the magnitude and the sign of intermolecular exchange.  Our theory should help to choose molecules with tailored exchange properties  when it comes to designing supramolecular structures that realize spin lattice models, in the spirit of the realization of the Haldane spin chain with $S=1$ triangulenes~\cite{mishra21b}.

Our analysis shows that a quantitative prediction of exchange interactions is very hard for two reasons. First, the convergence of CAS calculations as the dimension of the active space is increased is not obtained within our computational capabilities. Perturbation theory, discussed in Sec.~\ref{sec:full_pt}, shows a slow convergence of the contribution of higher-energy molecular orbitals, questioning the use of that indicator to truncate the Hilbert space. This calls for the use of other strategies, such as  density matrix renormalization group (DMRG) methods, to provide alternative methods to sample the Hilbert space~\cite{fano98}.
  
The second source of uncertainty is the strength of the Coulomb interaction in the experimentally relevant situation of molecules deposited on a substrate. A quantitative determination of screening would be needed. Our calculations show that going beyond the Hubbard approximation for the electron-electron interactions changes quantitatively the computed values of intermolecular exchange. We cannot rule out that in some limit, full-fledges long-range Coulomb interaction can lead to a failure of the OL rules. For instance, for graphene in the quantum Hall regime at half-filling, the Hubbard model predicts a $S=0$ GS whereas long-range Coulomb interactions predict a quantum Hall ferromagnet~\cite{nomura06}. Both, the issue of the Hilbert space truncation and the choice of the proper screened Coulomb interaction  deserve future attention. Comparison of theoretical prediction with experimental results is also made difficult by the renormalization of excitation energies due to Kondo interactions, studied by us in a previous work~\cite{jacob21}.

Importantly, our theory of Coulomb-driven superexchange interactions and the demonstration that these can be either ferromagnetic or AF provide a microscopic explanation of the physical mechanisms that enforce the OL rules~\cite{ovchinnikov78,Lieb89}. Note that our CDSE mechanism is analogous to the double spin polarization described for binuclear transition metal complexes~\cite{loth81,calzado02}.

Finally, the present work serves as a first step towards the study of intermolecular exchange for multi-radicals, such as $[n]$triangulenes with $n>2$.
It has been established that in the case of $S=1$ triangulenes the effective intermolecular spin coupling includes higher order terms such as
$(\vec{S}_1\cdot\vec{S}_2)^2$~\cite{mishra21b,catarina22,martinez-carracedo22}.
The approach of the present work applied to these systems should provide a microscopic understanding of their origin, as demonstrated already on the level of toy models~\cite{catarina22}.

\acknowledgments
We acknowledge fruitful discussions with R. Ortiz, G. Catarina and A. Costa.
We acknowledge financial support from the Ministry of Science and Innovation of Spain (Grants No. PID2020-112811GB-I00 and No. PID2019-109539GB-41), 
from Eusko Jaurlaritza (Grant No. IT1453-22), from Generalitat Valenciana (Grant No. Prometeo2021/017), from Fundacao Para a Ciencia e
a Tecnologia, Portugal (Grant No. PTDC/FIS-MAC/2045/2021), from the Swiss National Science Foundation,  Pimag grant, and from 
FEDER / Junta de Andaluc\'ia--Consejer\'ia de Transformaci\'on Econ\'omica, Industria, Conocimiento y Universidades
(Grant No. P18-FR-4834).

\appendix

\begin{figure}[b]
  \includegraphics[width=0.9\linewidth]{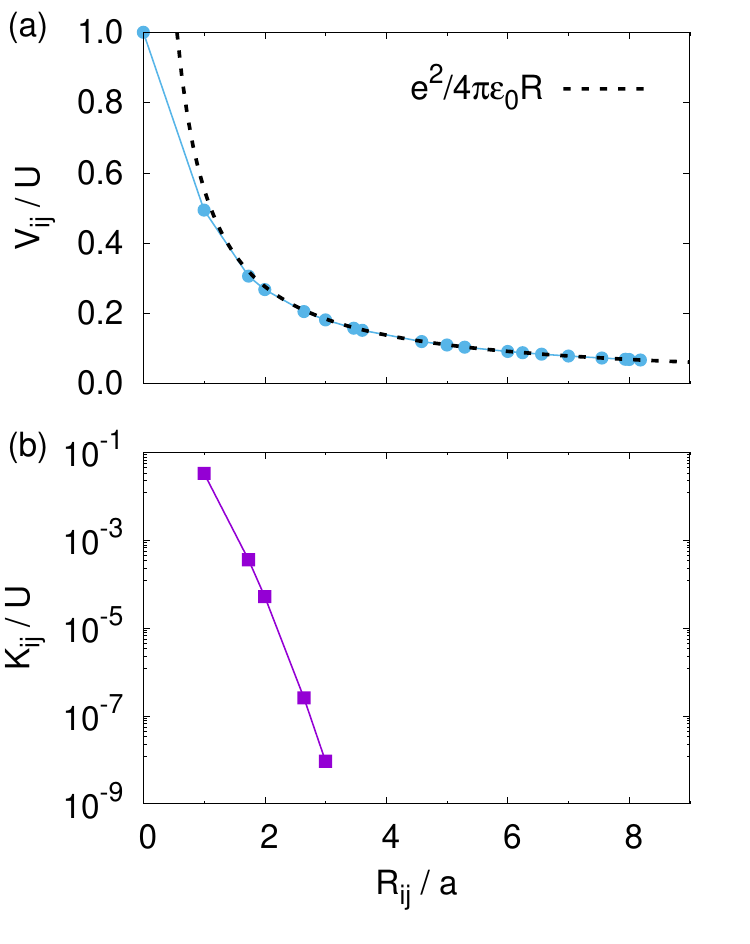} 
  \caption{ \label{fig:lr_coulomb}
    LR part of Coulomb interaction as a function of distance $R_{ij}$ between two carbon sites $i$ and $j$
    in units of the on-site (Hubbard) interaction $U$.
    (a) Coulomb repulsion $V_{i,j}\equiv\bra{\phi_i,\phi_j}\hat{v}_c\ket{\phi_i,\phi_j}$.
    (b) Direct Coulomb exchange $K_{i,j}\equiv\bra{\phi_i,\phi_j}\hat{v}_c\ket{\phi_j,\phi_i}$.
  }
\end{figure}

\section{LR Coulomb interaction}
\label{app:lr_coulomb}

In Fig.~\ref{fig:lr_coulomb} we show the LR part of the Coulomb interaction as a function of the distance
between carbon sites computed for the $p_z$ orbitals of the phenalenyl dimer (see below) using Gaussian09~\cite{G09}
with the LANL2MB minimal basis set. For distances larger than or equal to nearest neighbors the
Coulomb repulsion follows in good approximation the $1/r$ dependence of the bare Coulomb potential.
The direct  Coulomb exchange, on the other hand, decays exponentially with the distance, rendering all matrix $K_{i,j}$
elements beyond first neighbors ($R_{ij}>a$) negligible.

\bibliography{biblio}{}

\end{document}